# Multicriteria design and experimental verification of hybrid renewable energy systems. Application to electric vehicle charging stations

Paula Bastida-Molina[1*], Elías Hurtado-Pérez[1], María Cristina Moros Gómez[1], Carlos Vargas-Salgado[1]

[1] Instituto Universitario de Investigación en Ingeniería Energética, Universitat Politècnica de València, Valencia, 46022, Spain

[*] Corresponding author: paubasmo@etsid.upv.es

## Abstract

The installation of electric vehicle charging stations (EVCS) will be essential to promote the acceptance by the users of electric vehicles (EVs). However, if EVCS are exclusively supplied by the grid, negative impacts on its stability together with possible $CO_2$ emission increases could be produced. Introduction of hybrid renewable energy systems (HRES) for EVCS can cope with both drawbacks by reducing the load on the grid and generating clean electricity. This paper develops a methodology based on a weighted multicriteria process to design the most suitable configuration for HRES in EVCS. This methodology determines the local renewable resources and the EVCS electricity demand. Then, taking into account environmental, economic and technical aspects, it deduces the most adequate HRES design for the EVCS. Besides, an experimental stage to validate the design deduced from the multicriteria process is included. Therefore, the final design for the HRES in EVCS is supported not only by a complete numerical evaluation, but also by an experimental verification of the demand being fully covered. Methodology application to Valencia (Spain) proves that an off-grid HRES with solar PV, wind resources and batteries support would be the most suitable configuration for the system. This solution was also experimentally verified.

# 1. Introduction

By the end of the 20th century, climate change became one of the most disturbing global issues. The exorbitant amount of greenhouse gases (GHG), especially $CO_2$ emissions, sent to the atmosphere is leading to an environmental destruction, whose effects could be very detrimental for the nature and, as a consequence, for our society [1,2].

Transport sector has traditionally depended on fossil fuels, which are non-renewable resources and the main responsible for $CO_2$ emissions [3]. For instance, almost 93% of the global transport consumption in 2017 derived from oil products [4]. Moreover, around 23% of total $CO_2$ emissions in the world were generated by this sector [5]. Due to two different reasons: finite oil resources and environmental concerns, efforts have focused on the electrification of the transportation sector [6]. Hence, a high penetration of EVs is expected to happen in almost all developing countries in a short/mid-term future [7,8]. Despite the environmental suitability of these vehicles while riding on the roads, two drawbacks arise in this context. On the one hand, the extra electricity generated to cover EVs demand could lead to an increase of $CO_2$ emissions depending on the carbon intensity (CI) of the power sources involved in the electricity generation system [9,10]. On the other hand, this electricity increase could create negative impacts on the grid when recharging strategies remain unscheduled, concentrating the electrical consumption in peak demand hours [11–14]. The use of microgrids with integration of renewable sources to recharge EVs emerges as a solution to cope with the two previously mentioned difficulties [15]. First, the low CI of the renewable sources would decrease the $CO_2$ emissions generated during the electricity generation stage. Secondly, the pressure on the grid would decrease due to the demand reduction by using these microgrids [16]. These microgrids, known as Hybrid Renewable Energy Systems (HRES), combine the potential of different renewable sources: solar photovoltaic, wind generators, biomass gasifiers, etc., with the possibility to be supported by the grid or by other dispatchable resources as batteries, diesel generators or even hydrogen system in the most cutting-edge systems.

Currently, the number of electric vehicle charging stations (EVCS) is very limited and far enough to cope with the expected introduction of electric vehicles (EVs) in the coming years. In fact, the concerns of being unable to find an EVCS to recharge the EVs emerges as one of the highest barriers for the users to acquire this kind of vehicles [17]. Therefore, the development of fast recharging strategies together with the integration of renewable sources result essential to the integration and acceptance of EVs in our society. Several studies have addressed these topics. For instance, Huang et al. [18] developed a novel Geographic Information System to select the optimal location for the installation of new renewable EVCS depending on the current number of charging stations and renewable potentials, with the aim of minimizing the life cycle cost of the EVCS. Regarding the design process for the configuration of the HRES for EVCS, Domínguez-Navarro et al. [19] used a genetic algorithm to determine the HRES configuration for EVCS that maximizes the profit measured by its Net Present Cost (NPC), selecting finally a configuration with renewable generation and storage resources. Chowdhury et al. [20] studied the incorporation of a HRES for EVCS supported by the grid at the University Campus in Dhaka (Bangladesh), achieving a 21% of renewable generation and reducing GHG emissions by 52.9 t$CO_2$/year. They used software HOMER® [21] for the optimization process, looking for the lowest NPC configuration. Study [22] presents the configuration design process of an energy storage HRES in a rural community of the Democratic Republic of Congo with no access to the electrical grid for the recharge of electric Tuk-tuks (a traditional means of transport of the Democratic Republic of Congo). The installation of this HRES enhances the replacement of the traditional combustion engine Tuk-tuk vehicles by electric ones, together with the future deployment of EVs in these rural areas. Similarly, research in [23] boosts also the use of off-grid HRES systems for EVCS in rural remote areas. Namely, this research discusses the best configuration option for an EVCS in Labuhan Bajo (Indonesia) considering three types of batteries for energy storage: Lead Acid, Li-Ion (NCA) and Lithium Ferro Phosphate (LFP).

The methodologies presented in these above-mentioned studies only rely on economic parameters to design the HRES configuration for EVCS. However, other studies indicate that more parameters have to be considered for the system optimisation. For instance, Karmaker et al. [24] used also the HOMER ® code to decide the configuration of the HRES in an EVCS, but analized also the technical, economic and environmental feasibility of the selected configuration. Rashid et al. [25] focus the study on the electrical production and cost analysis, whereas Tulpule et al. [26] included environmental impacts, together with economic ones, in the design.

Another important issue to consider in the application of HRES to EVCS is the experimental validation of any optimized design. There are very few studies in this direction. In particular [27,28] state that, despite the suitability of numerical methodologies, the experimental verification of the HRES configuration ensures its reliability and real implementation. Research [27] describes the experimental results of a fast EVCS based on solar PV, wind sources and fuel cells and the necessity of implementing these systems in many remote regions of Russia with grid-connection problems. Research [28] focuses on the power system analyses of a microgrid that combines solar PV, utility grid and batteries to supply a fast charging EVCS. The experimental results verify the current flow and power balance of the system that were previously calculated with a simulation software.

Hence, this paper proposes a novel methodology that tries to cope with both aspects: to develop a weighted iterative multicriteria process based on economic, environmental and technical parameters to design the configuration of HRES in EVCS, and the experimental validation of the deduced designs by using a power balance and State of Charge (SOC) boundary criteria. The method is based on a previous characterization stage of the system in terms of energy by determination of the electricity demand of the EVCS and the evaluation of the local energy resources.

The study includes the application of the developed methodology, including the experimental verification, to Valencia (Spain). This region is expected to have a step mobility transition to EVs according to the Electric Mobility Plan [29], approved in 2007 by the Valencian Ministry of Sustainable Economy, Productive Sectors, Trade and Work. The plan aims to achieve an increasing penetration of both EVs and recharging points: 2030 EVs and 105/350 fast/semi-fast recharging points by the year 2020; 78.100 EVs and 210/950 fast/semi-fast recharging points in the year 2025 and 260.000 EVs and 270/2100 fast/semi-fast recharging points by the year 2030. Moreover, this Plan is framed within the Valencian Climate Change and Energy Strategy 2030 [30], which looks for the reduction of the GHG emissions, the inclusion of renewable sources in electricity generation and the energy efficiency enhancement by 2030. This legal framework boosts the installation of fast recharge points for the expected EVs fleet in the Comunidad Valenciana. Moreover, the renewable supply of such stations arises also as an environmental breakthrough to achieve, in line with that Strategy 2030. In this context, the application of the methodology presented in this paper for the design of a HRES for EVCS in the roads of Valencia has a remarkable interest.

The paper is organized as follows: section 2 presents the methodology, section 3 describes the case study of Valencia and section 4 provides the results and discussion of this application. Finally, the paper conclusions are outlined in section 5.

## 2. Methodology

This section presents the methodology developed to design a Hybrid Renewable Energy System to supply the electricity demand of EVCS. The method contemplates four different stages. The first one comprises the electricity demand modelling of the EVCS, together with the evaluation of the local energy resources analysis to determine the renewable technologies to be considered. The second phase makes an initial predesign of the system based on the NPC optimization by using the HOMER® software. Then, all the obtained configurations are evaluated and ranked in the third stage by using a multicriteria process that takes into account the technical, economic and environmental aspects for each of them. Finally, the last phase of the methodology addresses the experimental validation of the best-positioned configurations.

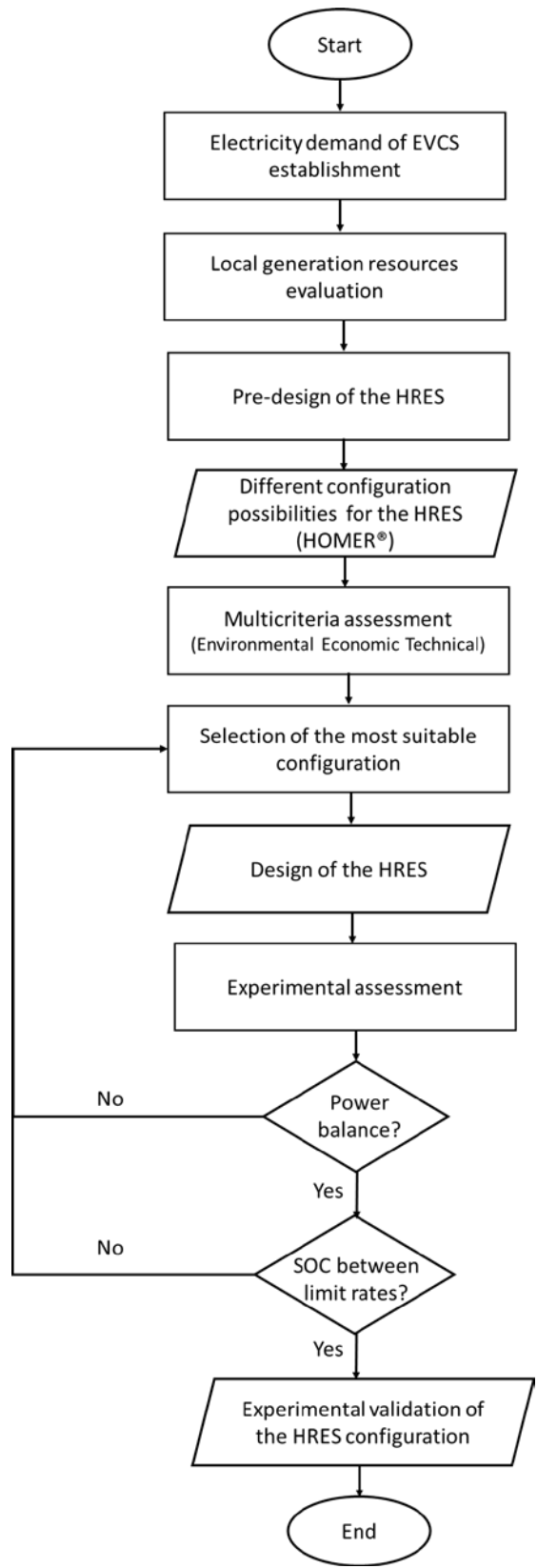

Figure 1. Flowchart of the proposed methodology.

## 2.1. Electricity demand of electric vehicles charging stations

EVCS demand depends on total amount of EVs refilling their batteries in the station and on the power consumption of each of these EVs. Regarding the first factor, this methodology establishes a temporary curve for each type of EV recharging in an EVCS: Battery Electric Vehicles (BEVs) and Plug-in-Hybrid Electric Vehicles (PHEVs), considering also their nature (cars and motorcycles). Taking a base fleet affected by two rates (penetration and recharge of EVs in the station [11]), the method determines each curve making use of eq.(1):

$$n(i,t) = N(t) \cdot f(i) \cdot r(i) \tag{1}$$

where n(i,t) is the number of EV of type i ( i=1 for BEV cars, i=2 for PHEV car and i=3 for BEV motorcycle) recharging at time t; N(t) represents the total number of vehicles on the road passing by the EVCS at that time, f(i) represents the fraction of these vehicles being electric and r(i) is the rate of those EVs needing recharge.

Referring to the second factor, the capacity of the battery, together with its state of charge (SOC) and the time duration of the recharging process determine the power demand of each EV type while recharging [24]. This power demand is given by eq.(2):

$$P_{EV}(i) = \frac{C_{bat}(i) \cdot [SOC_{Max} - SOC]}{T(i)} \tag{2}$$

Where $P_{EV}(i)$ corresponds to the power demand of EVs; $C_{bat}(i)$ represents the capacity of the EVs' batteries; $SOC_{Max}$ is the maximum level of the batteries' state of charge; $SOC$ corresponds to the real level of the batteries state of charge and $T(i)$ represents the time duration of the recharging process.

Finally, the power demand of the EVCS, $P_{EVCS}(t)$ arises as the electrical demand of every type of EV recharging there, like eq.(3) indicates:

$$P_{EVCS}(t) = \sum_{i} n(i,t) \cdot P_{EV}(i) \tag{3}$$

## 2.2. Local energy resources evaluation

At this stage, the methodology should determine the availability of renewable resources to be included in the HRES for EVCS. This implies the determination, for the place where the EVCS will be located and with the highest possible resolution, of some parameters: the solar irradiation and the clear index average [31], wind speed measured at the wind turbine height [32], the sustainable biomass production availability [33], etc.. Moreover, the necessity to support the system with batteries, the grid or with a generator should be also considered as potential back up to guarantee the reliability of the HRES in the EVCS.

## 2.3. Predesign of the HRES

HOMER® Pro software [21] is a well-known and widely used tool in the design of HRES, including its application to EVCS [23,25]. With the information of the technological options and the local resources to include in the HRES as an input to HOMER®, a list of different configurations for the system, ranked by their NPC, is obtained.

Despite the importance of the economic factor, the design of HRES for EVCS should also rely on environmental and technological criteria [24]. In line with this consideration, the present method utilizes the software HOMER® only in a predesigning phase of the HRES for the EVCS.

**2.4. Multicriteria assesment**

After the pre-design stage of the HRES, all the configuration options proposed by HOMER® are rank ordered using the methodology proposed in this section (2.4), based on a weighted multicriteria assessment of environmental, economic and technical parameters. This section describes the parameters and the multicriteria methodology.

**2.4.1. Environmental criteria**

The introduction of EVs is intended for a decarbonisation of the transport sector [5,34,35]. However, a recharge of these vehicles exclusively based on the grid could even lead to an increase of the emissions, depending on the carbon intensity (CI) generation mix of the grid [9,10,35]. Hence, this methodology proposes two factors to assess the environmental suitability of using a HRES for the EVs recharge in EVCS: $CO_2$ emissions reduction and renewable generation degree.

$CO_2$ emissions reduction (EmR)

This parameter determines the relative reduction in carbon emissions while using a HRES instead of the grid alone to supply the EVCS. $CO_2$ emissions reduction (EmR) can be obtained using eq. (4).

$$EmR = \frac{[E_{grid} \cdot g_{grid}] - [E_{HRES} \cdot g_{HRES}]}{[E_{grid} \cdot g_{grid}]} \qquad (4)$$

Where $E_{grid}$ is the electricity demanded to the grid if the EVCS has no any HRES support; $g_{grid}$ is the emissivity of the electricity from the grid; $E_{HRES}$ is the electricity provided to the EVCS from a HRES, and $g_{HRES}$ is the emissivity of the electricity from the HRES.

Specifically, the emissivity for the HRES ($g_{HRES}$) corresponds to a weighted combination of the generation resources of the system, which depend on their energy generation influence (eq. (5)).

$$g_{HRES} = \sum_j \frac{E_{HRES_j}}{E_{HRES}} \cdot g_j \qquad (5)$$

$$E_{HRES} = \Sigma_j E_{HRES_j}. \qquad (6)$$

Being $E_{HRES_j}$ the electricity provided by the component $j$ of the HRES and $g_j$ its specific emissivity.

Extreme values for EmR are 0 (no renewable sources in the HRES) and 1 (full renewable system without any $CO_2$ emission), as eq. (7) indicates:

$$EmR \in \{0,1\} \quad (7)$$

Renewable generation degree (ReG)

The contribution of renewable sources to the electricity consumption of the EVCS is another significant factor when analysing the environmental behaviour of the system [36]. Eq (8) determines this parameter (ReG), where not only the renewable contribution to the HRES take part, but also the renewable percentage of the electricity taken from the grid by the HRES.

$$ReG = \frac{\sum_r E_{HRES_r} + x_r \cdot E_{HRES_{grid}}}{E_{HRES}} \quad (8)$$

Being $E_{HRES_r}$ the electricity coming from the renewable source *r* of the HRES, $E_{HRES_{grid}}$ the electricity taken by the HRES from the grid and $x_r$ the fraction of renewable contribution in $E_{HRES_{grid}}$.

ReG values are in the interval 0 (when no renewable sources are involved in the HRES and in the electricity grid) and 1 (if all the electricity used by the HRES, including the grid, is generated with renewable sources), as eq. (9) reflects:

$$ReG \in \{0,1\} \quad (9)$$

### 2.4.2. Economic criteria

The importance of a thorough economic analysis for the design of the HRES EVCS appears in a wide range of researches [22,23,37]. In this methodology, the economic study uses the levelized cost of energy (LCOE). This is a widely used parameter to compare and evaluate different electricity generation procedures [38–40]. The LCOE indicates the average total cost of building and operating the corresponding energy system per unit of total electricity generated along its lifetime [41], as eq. (10) shows:

$$LCOE = \frac{\sum_j \sum_{t=1}^{t=n} \frac{(I_{tj} + O\&M_{tj} + F_{tj})}{(1+r)^t}}{\sum_{t=1}^{t=n} \frac{(E_{HRES_t})}{(1+r)^t}} \quad (10)$$

Where $I_{tj}$, $O\&M_{tj}$ and $F_{tj}$ represent the investment cost, operation and maintenance cost and fuel cost respectively of each generation resource j at the time t into consideration of the lifetime of the system (n), whereas $r$ corresponds to the discount rate.

The methodology introduces a normalized LCOE ($NLCOE$) to compare the LCOE for an EVCS supplied by the grid ($LCOE_{grid}$) with the LCOE for an EVCS supplied by the HRES in study ($LCOE_{HRES}$), as eq. (11) indicates:

$$NLCOE = \frac{LCOE_{grid}}{LCOE_{HRES}} \tag{11}$$

Hence, an economic factor (EcF) for the multicriteria analysis can be defined as:

$$EcF = Min\ (1;\ NLCOE) \tag{12}$$

Moreover, EcF values range in the interval of 0 (for very high $LCOE_{HRES}$) and 1 (if the HRES has a lower LCOE that the grid one), as eq. (13) shows:

$$EcF \in \{0,1\} \tag{13}$$

### 2.4.3. Technical criteria

The technical study comprises two remarkable parameters: the security of supply and the adequacy sizing of the system.

Security of supply (SS)

This factor evaluates the guarantee of electricity supply taking into account the different combination of generation sources and back-up systems in the HRES for EVCS [40], as eq. (14) indicates.

$$SS = 1 - \sum_j (1 - f_j) \tag{14}$$

Being $f_j$ the reliability of the generation source $j$.

For non-dispatchable generation sources, i.e.: solar PV and wind generation, we can consider the magnitude of the energy contribution related to the demanded one and the fraction of the time these sources are available, as eq. (15) indicates.

$$f_j = Min\left[1;\ \frac{E_j}{E_{EVCS}}\right] \cdot \delta_j \tag{15}$$

Where $E_j$ represents the electricity provided by the non-dispatchable sources in question, $E_{EVCS}$ is the total electricity demanded by the EVCS and $\delta_j$ corresponds to the fraction of hours that the source is available.

For dispatchable electricity sources, such as the grid and the backup generator, eq. (16) determines their feasibility as follows:

$$f_j = Min\left[1; \frac{P_j}{P_{EVCS}}\right] \cdot \delta_j \qquad (16)$$

Where $P_j$ represents the generator maximum power and the contracted power from the grid, and $P_{EVCS}$ corresponds to the maximum power of the EVCS. Values for the security factor $\delta_j$ are available for diesel generators [42] and for the grid [43,44].

In the case of the storage battery bank, the feasibility factor can be defined as:

$$f_b = Min\left[1; \frac{E_b}{E_{EVCS}}\right] \cdot \delta_b \qquad (17)$$

Being $E_b$ the nominal capacity of the battery bank and $\delta_b$ the security factor, also available in [45].

SS values are in the interval of 0 (when the system cannot ensure the electricity supply at all) and 1 (if the security of supply is completely assured), as eq. (18) reflects:

$$SS \in \{0,1\} \qquad (18)$$

Electricity sizing adequacy (ESA)

Finally, this last parameter assesses the adequacy of the system in relation to its power sizing. Systems should be designed in such a way that they cover all the demand requirements, but the minimum excess of generation, as eq. (19) indicates.

$$ESA = Min\left[1; \frac{E_{EVCS}}{E_{HRES}}\right] \qquad (19)$$

ESA values are in the interval of 0 (when the power sizing is not adequate at all) and 1 (If its power sizing is completely achieved), as eq. (20) indicates:

$$ESA \in \{0,1\} \qquad (20)$$

### 2.4.4. Multicriteria assessment

In this stage, the proposed methodology evaluates all the configuration possibilities obtained in the predesigning phase with HOMER® Pro Software for the HRES EVCS in question. For this evaluation, the methodology applies a weighted multicriteria assessment on each of these configurations based on the above-explained criteria. Hence, a merit figure (CP) is deduced for each configuration option.

Table 1 shows the evaluation criteria together with their corresponding weighting factors. Moreover, eq. (21) describes the multicriteria evaluation for each configuration, where constraint (22) applies.

Table 1. Criteria and weighting factors for the evaluation.

|  | Criteria | Weighting factor |
|---|---|---|
| Environmental | CO$_2$ emissions reduction (EmR) | α$_{EmR}$ |
|  | Renewable generation degree (ReG) | α$_{ReG}$ |
| Economic | Economic Factor (EcF) | α$_{EcF}$ |
| Technologic | Security of supply (SS) | α$_{SS}$ |
|  | Electricity sizing adequacy (ESA) | α$_{ESA}$ |

$$CP = \alpha_{EmR} \cdot EmR + \alpha_{ReG} \cdot ReG + \alpha_{EcF} \cdot EcF + \alpha_{SS} \cdot SS + \alpha_{ESA} \cdot ESA \qquad (21)$$

$$\alpha_{EmR} + \alpha_{RG} + \alpha_{EcF} + \alpha_{SS} + \alpha_{ESA} = 1 \qquad (22)$$

Finally, once all the configurations have been analyzed, they are ranked in accordance with their CP values. Hence, the one with the highest value would be the best design solution for a HRES in an EVCS, based on a complete study of the system including environmental, economic and technical aspects.

### 2.5. Experimental verification of the hybrid renewable energy system

The last stage of the methodology consists of an experimental verification of the selected design for the HRES in the EVCS after the previously explained multicriteria assessment phase [28,46]. The theoretical design needs to be accurately reproduced in a laboratory with all the required technologies. Therefore, a scaled version of the selected configuration results necessary [36], being the scale factor (SF) determined by the capabilities of the experimental system to be used ($P_{lab}$), and the maximum power of the EVCS ($P_{EVCS}$) as eq. (23) indicates:

$$SF = \frac{P_{EVCS}}{P_{lab}} \qquad (23)$$

Consequently, this scale factor affects the EVCS power demand curve, determined in section 2.1., so that the experimental EVCS power demand $(P_{EVCS\,exp}(t))$ is determined by eq. (24). The power of each generation system $(P_j)$ is scaled as well, being the experimental generation power $(P_{j\,exp})$ obtained by eq.(25).

$$P_{EVCS\,exp}(t) = \frac{P_{EVCS}(t)}{SF} \tag{24}$$

$$P_{j\,exp} = \frac{P_j}{SF} \tag{25}$$

The methodology imposes two conditions to be satisfied before accepting the system configuration [32,36,45]. Firstly, the EVCS load requirements should be covered at each time of the day, so to reach this goal the power balance should accept a certain rate of power losses ($L$) in the system (eq. (26)). Then, for systems with a storage capacity based on batteries, the state of charge (SOC) of these batteries should be all the time in the range between the allowed minimum and maximum values. (eq. (27)).

$$\frac{|\sum P_{j\,exp}(t) - P_{EVCS\,exp}(t)|}{P_{j\,exp}(t)} \leq L \tag{26}$$

$$SOC_{min} \leq SOC(t) \leq SOC_{max} \tag{27}$$

The fulfilment of these conditions ensures the correct design of the HRES for the EVCS. If any of them were not met, the methodology includes an iterative process on the selection of the theoretical design of the system, following the rank order deduced from the multicriteria assessment.

## 3. Case study: Valencia (Spain)

The paper applies the previously explained methodology to Valencia, the capital province of the Comunidad Valenciana, located in the East of Spain. This region is experimenting a deep ecological transition in terms of mobility motivated by its Electric Mobility Plan [29]. The plan establishes as final 2030 objective that the EVs represent 25% of the market share of the Comunidad Valenciana along with establishing one fast recharge point for every ten EVs. The final achievement of these goals would lead to a considerable GHG emission reduction, specifically the Plan foresees a total 622.000 tons of $CO_2$ emissions decrease. This Plan is framed within the Valencian Climate Change and Energy Strategy 2030 [30], whose three central goals lie in a reduction of the GHG emissions, the renewable sources increase in electricity generation and a substantial energy efficiency enhancement by 2030.

This legal framework boosts the installation of fast recharge points for the expected EVs fleet in the Comunidad Valenciana, namely in the form of EVCS. Moreover, the renewable supply of such stations arises also as an environmental breakthrough to achieve, in line with the above mentioned 2030 Energy Strategy. In this context, we decided to apply the methodology presented in this paper to the design of a HRES for EVCS in the roads of Valencia in an imminent future. Moreover, this work only considers the recharge of light electric vehicles (LEVs) in EVCS with possibilities of recharge: BEV cars, PHEV cars and BEV motorcycles. Nowadays, heavy internal combustion vehicles, like private buses or trucks, represent a 15% of the recharge in petrol stations located at roads of Valencia [47]. However, the currently available batteries of their equivalent heavy EVs are not yet developed enough to provide the autonomy desired by these vehicles in roads [11]. Therefore, it is not realistic to assume this type of vehicles being recharged at EVCS.

### 3.1. Electricity demand of electric vehicles charging stations

The determination of the EVCS electricity demand in Valencia could be deduced from the current flow of light internal combustion engine vehicles (LICEVs) passing by the roads close to petrol stations in the region. The accurate traffic information for Valencian territory provided by the Spanish data base [47] allowed us to model the average flow of LICEVs in study, represented by N(t) in eq. (1) (Figure 2). 94% of this fleet was formed by cars and 6 % by motorcycles.

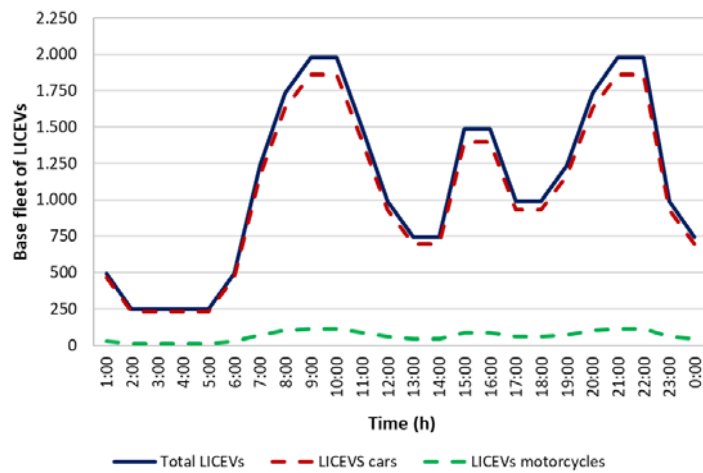

Figure 2. Base fleet of LICEVs for Valencia N(t).

The rate of penetration of LEVs in this base fleet of LICEVs will match their expected penetration in the Spanish fleet by an imminent future [35]: 2.5% for BEVs cars, 2.5% for PHEVs cars and 5% for BEVs motorcycles, considering just the LEVs with possibilities of recharging in EVCS due to their configuration (BEVs and PHEVs) [48].

Finally, study [49] claims that the percentage of LEVs passing by the EVCS that will finally recharge there is expected to be slightly higher than the equivalent traditional refueling behavior. Hence, this percentage increases up to 6%. Table 2 reflects all these parameters to be used in eq. (1).

Table 2. Rate of penetration and recharge of LEVs.

|  | f (%) | r (%) |
|---|---|---|
| BEVS cars | 2.5 | 6 |
| PHEVS cars | 2.5 | 6 |
| BEVS motorcycles | 5 | 6 |

For the determination of the power consumption of each type of LEV at EVCs, we made a detailed analysis on their battery capacity, SOC and required time for recharging at the EVCS, assuming only a fast recharging mode [11,50]. Regarding the first parameter, researches [51–53] shed light on the determination of battery capacity for BEVs cars and motorcycles, and PHEVs cars. Referring to the initial SOC, we took the hypothesis that the SOC for the LEVs recharging at the EVCS will be 20% [54]. Table 3 indicates the assumed values for the different parameters of the full recharge of the different types of EV.

Table 3. LEVs' recharging parameters

|  | $C_{bat}$ (kWh) | $SOC_{Max}$ (%) | SOC (%) | T (min) | $P_{EV}$ (kW) |
|---|---|---|---|---|---|
| BEVs cars | 40 | 100 | 20 | 40 | 48 |
| PHEVs cars | 14 | 100 | 20 | 14 | 48 |
| BEVs motorcycles | 3 | 100 | 20 | 3 | 48 |

Using these data, it is possible to deduce the electricity demand of the EVCS for the Valencian case study, shown at Figure 3. Maximum power demand is 270 kW, and takes place during the early morning (from 9:00 to 10:00) and at early night again (from 21:00 to 22:00). Final contribution of BEVs motorcycles to the electricity demand results in a 6%, in front of 49% for BEVs cars and 45% for PHEVs cars, respectively.

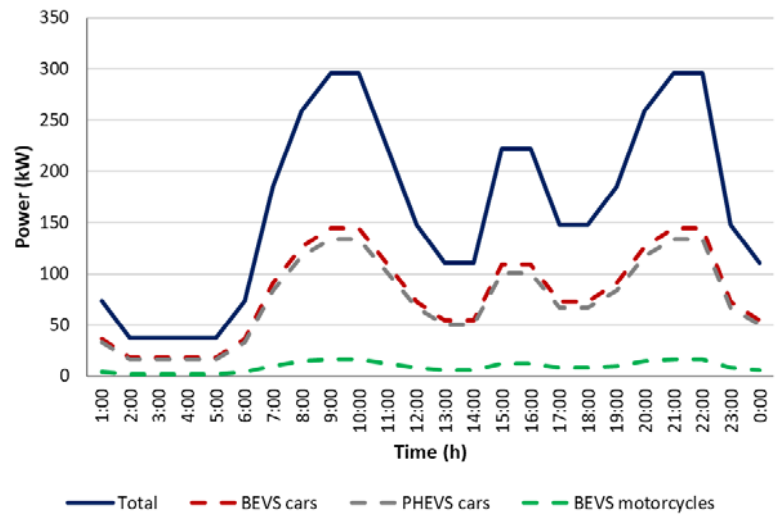

Figure 3. Electricity demand in EVCS.

## 3.2. Generation resources analyses

Valencia is a province located at the East of Spain, next to the Mediterranean Sea. Its geographical position corresponds to the coordinates 39°28'00"North 0°22'30"West and it has an elevation of 16 meter above sea level. The analysis of the renewable potential of Valencia highlighted solar resources as the most suitable ones, followed by wind resources.

According to PVGIS-CMSAF [55], Valencia has an average annual irradiation of 1735 kWh/m$^2$/year with the monthly dependence shown at Figure 4. The highest irradiation data corresponds to the summer months, reaching the highest values in June and July, with approximately 7.8 kWh/m$^2$/day. On the contrary, the lowest irradiation values correspond to the winter months, specifically December and January, with 2.1 and 2.5 kWh/m$^2$/day respectively. From these data, we can deduce an average solar daily irradiation of 5 kWh/m$^2$/day and a clearness average index of 0.65.

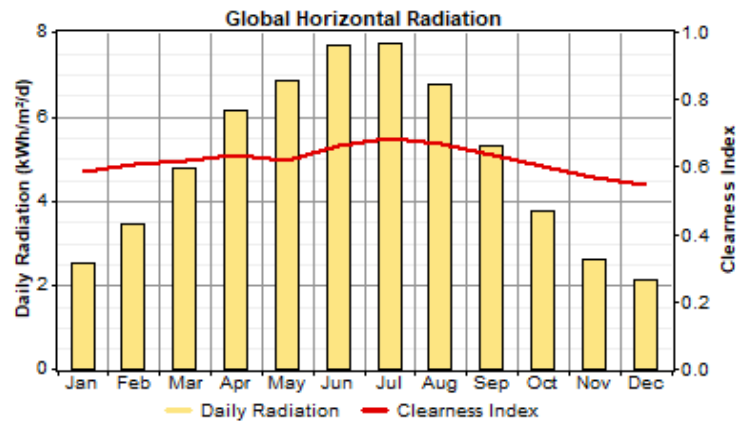

Figure 4. Average solar daily irradiation and clearness index in Valencia.

Moreover, data from [56] indicated that the average wind speed of Valencia is 3.6 m/s, measured at 18 m above the ground. Figure 5 reflects the daily average data for each month. These values revealed the suitability of wind resources in Valencia, although they do not have the high potential of the solar resources. The availability of this resource presents a trend which results ideal for the HRES: solar irradiation offers its highest values during summer months; meanwhile wind speed reaches the highest data during the winter ones. Hence, each type of renewable generation would ideally complement the other, helping to the reliability of the HRES.

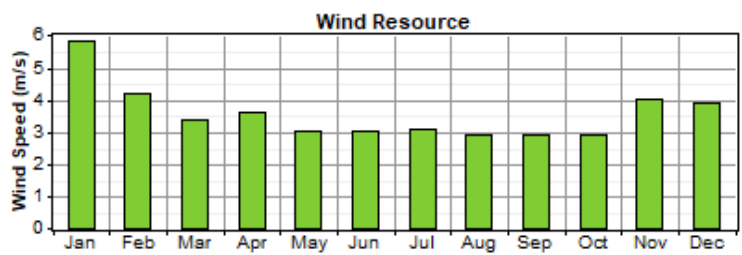

Figure 5. Wind speed in Valencia.

Regarding back-up systems, grid connection is a feasible possibility for EVCS [25], since Valencia is a complete electrified area. Furthermore, batteries and diesel generators can be also considered as possibilities to support the HRES, especially if the EVCS is intended to be off-grid [22].

### 3.3. Inputs for the design of the hybrid renewable energy system

Taking into account the power demand from the EVCS and the availability of solar and wind resources in Valencia, an initial estimation of the HRES system configuration to be used as input for the HOMER simulation was defined (Table 4).

Table 4. HRES EVCS components sizing.

| Solar PV (kW) | Wind (kW) | Grid connection (kW) | Diesel Generator (kW) | Battery (kWh) |
|---|---|---|---|---|
| 500 | 330 | 270 | 280 | 960, 1920, 2880, 4800 |

To ensure a reliable supply, the maximum acceptable capacity shortage of the system was established in 10% for the HOMER® simulations. HOMER ® results provided a list with 55 configuration possibilities ordered by their NPC values. Before applying the multicriteria evaluation, configurations without renewable generation were discarded. Besides, alternatives including grid and diesel generator were also rejected, considering the generator was not necessary in the presence of grid. Table 5 summarizes the discarded design options, meanwhile Table 6 reflects the 27 selected configuration alternatives to be analysed with the multicriteria methodology.

Table 5. Discarded design options.

| Discarded design scenarios | HOMER options | Reason |
|---|---|---|
| Grid | 2 | |
| Grid + gen | 5 | |
| Grid + bat | 13, 22, 28, 36 | |
| Grid + gen + bat | 17, 24, 31, 40 | Lack of renewable generation. |
| Gen + bat | 50, 51 | |
| Gen | 55 | |
| Ren + grid + gen | 3, 7, 14 | The diesel generator does not contribute to energy generation, due to the presence of the grid. |
| Ren + grid + gen + bat | 8, 12, 19, 20, 21,26 27, 33, 34, 35, 41,42 | |

gen: diesel generator; bat: batteries; ren: renewable resources.

Table 6. Selected configuration options to be analysed by the methodology.

| | HOMER Option | Solar PV (kW) | Wind (kW) | Grid connection | Generator (kW) | Battery (kWh) |
|---|---|---|---|---|---|---|
| Ren + grid | 1 | 500 | 0 | Yes | 0 | 0 |
| Ren + grid | 4 | 0 | 330 | Yes | 0 | 0 |
| Ren + grid + bat | 6 | 500 | 0 | Yes | 0 | 960 |
| Ren + grid + bat | 9 | 500 | 0 | Yes | 0 | 1920 |
| Ren + bat | 10 | 500 | 330 | No | 0 | 4800 |
| Ren + grid | 11 | 500 | 330 | Yes | 0 | 0 |
| Ren + grid + bat | 15 | 500 | 0 | Yes | 0 | 2880 |
| Ren + grid + bat | 16 | 0 | 330 | Yes | 0 | 960 |
| Ren + grid + bat | 18 | 500 | 330 | Yes | 0 | 960 |
| Ren + grid + bat | 23 | 500 | 0 | Yes | 0 | 1920 |
| Ren + grid + bat | 25 | 500 | 330 | Yes | 0 | 1920 |
| Ren + grid + bat | 29 | 500 | 0 | Yes | 0 | 4800 |
| Ren + grid + bat | 30 | 0 | 330 | Yes | 0 | 2880 |
| Ren + grid + bat | 32 | 500 | 330 | Yes | 0 | 2880 |
| Ren + gen + bat | 37 | 500 | 330 | No | 280 | 4800 |
| Ren + grid + bat | 38 | 0 | 330 | Yes | 0 | 4800 |
| Ren + grid + bat | 39 | 500 | 330 | Yes | 0 | 4800 |
| Ren + gen + bat | 43 | 500 | 330 | No | 280 | 2880 |
| Ren + gen + bat | 44 | 500 | 330 | No | 280 | 1920 |
| Ren + gen + bat | 45 | 500 | 0 | No | 280 | 4800 |
| Ren + gen + bat | 46 | 500 | 0 | No | 280 | 2880 |
| Ren + gen + bat | 47 | 0 | 330 | No | 280 | 2880 |
| Ren + gen + bat | 48 | 0 | 330 | No | 280 | 4800 |
| Ren + gen + bat | 49 | 0 | 300 | No | 280 | 1920 |
| Ren + gen | 52 | 500 | 330 | No | 280 | 0 |
| Ren + gen | 53 | 500 | 0 | No | 280 | 0 |
| Ren + gen | 54 | 0 | 330 | No | 280 | 0 |

The application of the multicriteria methodology to the Valencian case study required the definition of some input parameters regarding the environmental, economic and technical criteria, as well as the weighting factors.

Environmental criteria

The relative decrease of $CO_2$ emissions achieved when using a HRES instead of the traditional grid for charging vehicles in EVCS together with the renewable generation degree comprise the environmental factors to assess each design option for the system. Thus, the emissivity for each renewable source results of utmost importance, as well as the emissivity and renewable presence for the Spanish grid. A wide study of renewable and non-renewable sources' emissivity is available on [24] and [35,57] contain all the information regarding the Spanish electricity mix. Using this information, Table 7 summarizes the emissivity values to be used.

Table 7. Emissivity for generation sources and renewable contribution to the grid.

| | Solar PV | Wind | Diesel | Spanish grid |
|---|---|---|---|---|
| g (g $CO_2$/kWh) | 40 | 20 | 600 | 318.1 |
| $X_r$ (%) | - | - | - | 27.1 |

Economic criteria

This paper uses the NLCOE to assess the economic behaviour of each design option, where the economic modelling of such parameter includes the investment, operation and maintenance and fuel costs for each element of the HRES, as well as its corresponding discount rate and the lifetime of the project. A thorough research [32,36] was made to accurately determine these values for the case study. These are presented in Table 8.

Table 8. Economic modelling.

|  | Investment cost (€/kW) | O&M cost (€/kW) | Fuel cost (€/L) | n (years) | r (%) |
|---|---|---|---|---|---|
| Solar PV | 1200 | 40 | - | - | - |
| Wind | 2020 | 60 | - | - | - |
| Diesel generator | 380 | 1.5[a] | 1.05 | - | - |
| Batteries | 950[c] | 10[c] | - | - | - |
| Grid | - | 0.15[b] | - | - | - |
| General project | - | - | - | 25 | 8 |

[a]€/h; [b]Grid power price; [c]€/unit

Technical criteria

The technical evaluation of the methodology includes an analysis of the power selected for each power source together with the application of a security coefficient to each one of them to ensure the feasibility of the system. To determine this security coefficient for dispatchable technologies, study [42] quantifies its value for diesel generator, and [43,44] for the Spanish grid. Moreover, the security coefficient for batteries matches its depth of discharge according to [21]. This coefficient varies for non-dispatchable sources, depending on the number of equivalent hours (1735 for solar PV [55] and 1889 for wind in Valencia [56]). Table 9 summarises the security coefficient data for each generation source in the HRES.

Table 9. Security coefficient for the generation sources ($\delta_j$).

| Solar PV (%) | Wind (%) | Diesel generator (%) | Spanish Grid (%) | Batteries (%) |
|---|---|---|---|---|
| 19.8 | 21.6 | 85.7 | 98 | 70 |

Multicriteria assessment

The methodology presented in this paper allows users to arbitrarily decide through a series of weighting factors the importance that each criteria will play during the evaluation process. In this case of study, we have decided to apply a balanced evaluation process, where all the criteria have the same weight: 20% each.

## 4. Results and discussion

This section presents the results for the application of the methodology to the Valencian case study. Namely, it exposes the selected designs of the HRES in EVCS of Valencia after applying the multicriteria assessment, together with the experimental validation of such designs in the Laboratory of Distributed Energy Resources (LabDER) of the Polytechnic University of Valencia (UPV) [46].

### 4.1. Design of the hybrid renewable energy system: multicriteria assessment

The application of the multicriteria methodology presented in this paper to the Valencian case study gave rise to a rank ordered list of the design options for the HRES in EVCS. Table 10 reflects the individual percentage assessment of the environmental, economic and technical criteria for each option, as well as the final evaluation considering equal ponderation values for all of them.

Table 10. Multicriteria assessment of the HRES configurations. Selected designs for the HRES in EVCS.

| Configuration | HOMER position # | Methodology position # | EmR (%) | ReG (%) | EcF (%) | SS (%) | ESA (%) | Total (%) |
|---|---|---|---|---|---|---|---|---|
| Ren + bat | 10 | 1 | 88,84 | 100 | 83,13 | 83,29 | 88,85 | 88,82 |
| Ren + gen + bat | 37 | 2 | 67,95 | 91,04 | 68,56 | 98,14 | 80,89 | 81,32 |
| Ren + grid | 11 | 3 | 49,05 | 80,96 | 88,08 | 98,44 | 65,64 | 76,43 |
| Ren + gen + bat | 43 | 4 | 56,65 | 86,83 | 63,94 | 96,17 | 77,15 | 76,15 |
| Ren + grid + bat | 18 | 5 | 49,05 | 80,96 | 83,13 | 98,73 | 65,65 | 75,50 |
| Ren + grid | 4 | 6 | 31,70 | 57,81 | 97,79 | 98,20 | 88,62 | 74,83 |
| Ren + grid | 1 | 7 | 31,11 | 64,80 | 100,00 | 98,26 | 79,53 | 74,74 |
| Ren + grid + bat | 25 | 8 | 49,09 | 80,97 | 78,24 | 99,02 | 65,66 | 74,59 |
| Ren + grid + bat | 6 | 9 | 31,12 | 64,80 | 95,68 | 98,58 | 79,54 | 73,94 |
| Ren + grid + bat | 32 | 10 | 49,11 | 80,98 | 74,30 | 99,31 | 65,67 | 73,87 |
| Ren + grid + bat | 39 | 11 | 49,67 | 81,18 | 67,86 | 99,67 | 65,91 | 72,86 |
| Ren + grid + bat | 9 | 12 | 31,12 | 64,80 | 89,86 | 98,91 | 79,53 | 72,84 |
| Ren + grid + bat | 16 | 13 | 31,71 | 57,81 | 83,65 | 98,54 | 88,63 | 72,07 |
| Ren + grid + bat | 15 | 14 | 31,12 | 64,80 | 84,18 | 99,05 | 79,54 | 71,74 |
| Ren + grid + bat | 23 | 15 | 31,74 | 57,82 | 78,70 | 98,80 | 88,65 | 71,14 |
| Ren + grid + bat | 30 | 16 | 31,76 | 57,82 | 75,14 | 98,80 | 88,67 | 70,44 |
| Ren + grid + bat | 29 | 17 | 31,13 | 64,81 | 75,57 | 99,05 | 79,55 | 70,02 |
| Ren + gen + bat | 44 | 18 | 40,15 | 81,35 | 56,60 | 94,55 | 72,28 | 68,98 |
| Ren + grid + bat | 38 | 19 | 31,76 | 57,82 | 67,86 | 98,80 | 88,67 | 68,98 |
| Ren + gen + bat | 45 | 20 | 0 | 56,46 | 47,16 | 94,71 | 86,84 | 57,04 |
| Ren + gen + bat | 46 | 21 | 0 | 54,94 | 45,70 | 94,71 | 84,50 | 55,97 |
| Ren + gen + bat | 47 | 22 | 0 | 40,97 | 39,00 | 93,33 | 86,19 | 51,90 |
| Ren + gen + bat | 48 | 23 | 0 | 41,11 | 38,55 | 93,33 | 86,49 | 51,89 |
| Ren + gen + bat | 49 | 24 | 0 | 40,78 | 38,55 | 93,33 | 85,81 | 51,69 |
| Ren + gen | 52 | 25 | 0 | 59,77 | 25,63 | 91,30 | 53,11 | 45,96 |
| Ren + gen | 53 | 26 | 0 | 40,47 | 23,54 | 90,30 | 62,24 | 43,31 |
| Ren + gen | 54 | 27 | 0 | 31,69 | 22,06 | 90,01 | 66,67 | 42,09 |

*Note: the dimension values (kW or kWh) of each option could be consulted in Table 6.*

It is possible to see the difference between the method hereby presented and the one followed by HOMER® when assessing the alternatives. For instance, the best-valued option of this method corresponds to the 10[th] one of the HOMER® ranking, whereas the best-valued option using HOMER® corresponds to the 7[th] one of the multicriteria method. These outcomes result coherent with the behavior of both tools and verify one of the aims of the work: meanwhile HOMER® bases its evaluation just on an NPC optimization, the method hereby presented takes into account every factor that could affect HRES in EVCS, resulting in a more complete and realist evaluation.

Figure 6 shows the evaluation of each of the multicriteria parameters for each of the analyzed configurations. Regarding environmental parameters (EmR and ReG), the configurations with renewable generation and batteries are by far the most influential one. The configurations that include renewable generation, batteries and the support of diesel generators result also influential in environmental criteria for the options that use diesel generator during short periods. However, the design options that use diesel generators during long time periods have the worst environmental behavior. Alternatives including renewable generation with the support of the grid are the best economic options (EcF), and they also present good technical criteria (SS, ESA). However, configurations with renewable generation and diesel generators result the worst choice in all the aspects: environmental, economic and technical.

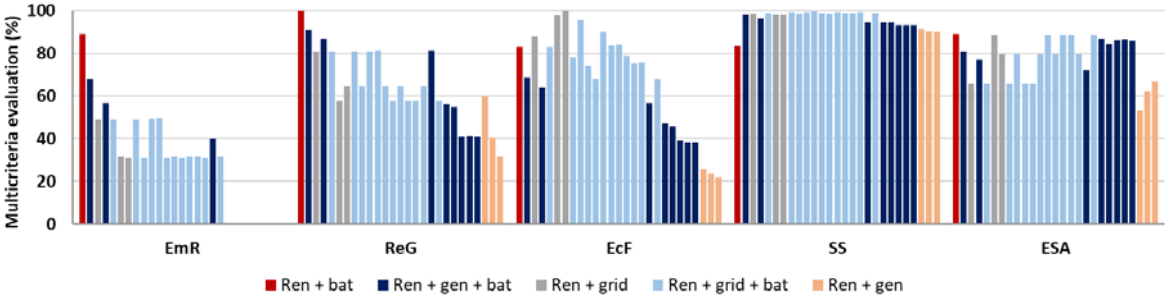

Figure 6. Multicriteria assessment.

*Note: the design options are ordered according to Table 10-Multicriteria methodology.*

From the configuration ranking in Table 10 we can deduce the three most suitable configuration options for the HRES in the Valencia case study. The highest-scored option is related to an off grid energy scenario that includes renewable generation (500 kW solar PV and 330 kW wind) and the support of a group of batteries (4800 kWh). The second alternative corresponds to another off grid scenario, similar to the first one, but with the support of a diesel generator (280 kW). The third-highest scored option finally represents an on-grid scenario, where the grid supports the renewable generation (500 kW solar PV and 330 kW wind).

Most of the pioneers HRES EVCS' projects developed in regions where grid connection results possible tend to rely on such kind of support for the system [24,58] mainly motivated by its ease of use, security of supply and economic performance. However, the multicriteria assessment presented in this paper reveals the influence that the environmental aspects could play in the selection process favoring off grid solutions, if possible. Figure 7 presents a comparison among the parameter evaluation for each of the three most suitable scenarios.

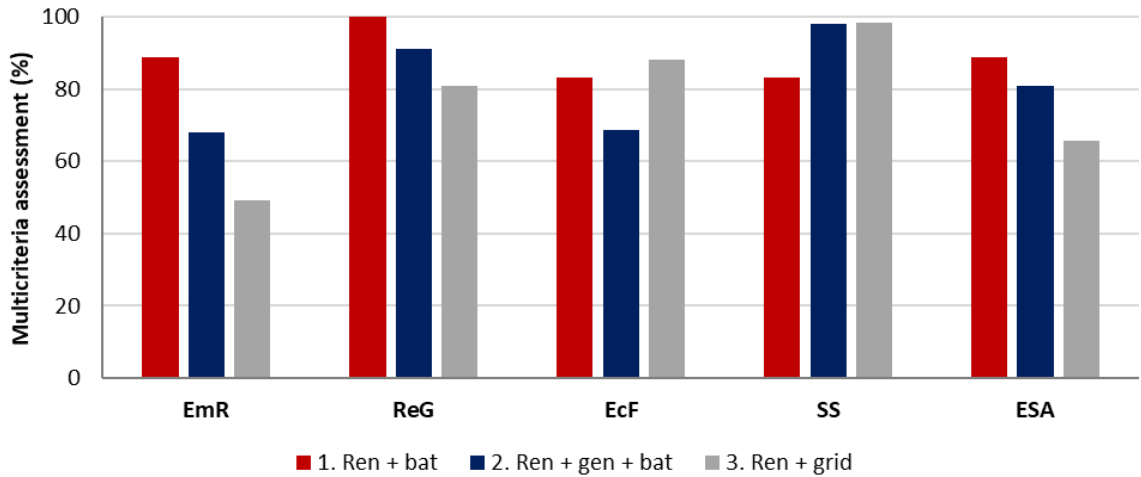

Figure 7. Selected designs for the HRES in EVCS.

*Note: the design options are ordered according to Table 10-Multicriteria methodology.*

The off grid configuration with renewable generation and batteries storage presents the best environmental behavior, since it does not depend on polluting sources. However, the second off grid configuration (renewable generation with diesel generator and batteries) is penalized by the use of the diesel generator. Moreover, the on-grid configuration, given the dependence of the Spanish electrical mix in some high polluting sources [57], is the worst in terms of environmental influence, especially referring to $CO_2$ reduction. However, this on-grid configuration arises as the most economic one, having the second off-grid configuration the lowest economic parameter due to the expenses of the diesel generator and its fuel. On the contrary, the on grid configuration together with the off grid configuration that includes a diesel generator have the highest security of supply, since they both count with dispatchable support sources.

## 4.2. Experimental verification of the hybrid renewable energy system

To conclude the complete design process of the HRES for EVCS for the case study, the selected design alternatives through the multicriteria assessment were experimentally validated in the Laboratory of Distributed Energy Resources (labDER) [46] of the Institute for Energy Engineering of the Polytechnic University of Valencia (Spain). This laboratory includes a hybrid combination of generation resources (2 kW$_p$ solar PV, 1.5 kW wind turbine, 10 kW biomass gasifier, 1.7 kW diesel generator, optimal grid connection and 1.2 kW fuel cell). It also includes storage systems (12 kWh batteries and 7 kW hydrogen system) and a programmable load system (from 0.5 to 9.2 kW) that enables to simulate any time dependence of the demand. Hence, to experimentally validate the suitability of the selected HRES configurations, they were reproduced in labDER with a scale factor of 1:250.

Each scaled experiment comprise a complete day of simulation for the three most suitable HRES designs for EVCS. For each simulation, the batteries SOC limits were fixed to 30% and 100%, according to their discharge limits. Moreover, authors fixed the maximum acceptable rate of power losses in 5%, considering previous experimental studies in such field [36,46].

Figure 8 plots the energy balance and SOC results for the highest-scored configuration, which includes renewable generation and the support of batteries.

At the beginning of the experiment, the demand requirements were the highest. However, at that period, solar irradiation was still low and wind contribution was practically null. Therefore, batteries contributed in part to meet electricity demand. Later, solar PV and wind contribution reached their maximum values. Hence, the HRES was able to meet the EVCS supply with an excess of energy, which was used to recharge batteries. The SOC of batteries increased during this period, achieving its full charge status. The highly fluctuating behavior of the wind generation, characteristic in small wind turbines like the labDER one [46], is also reflected in the power supplied by the batteries and in their SOC. At late afternoon, solar irradiation started to disappear and wind contribution was low. Finally, at night, both solar and wind contribution were null and load supply was based exclusively on batteries, reaching their lowest SOC value at the experiment in the early morning, when solar irradiation was again available and recharge was initiated again.

These results demonstrated the energy achieved with the HRES in question could cope with the assumed electricity demand. Moreover, the maximum rate of power losses in this experiment was 4.5% and the rates of batteries SOC oscillated between 35% and 100%. Hence, the experiment met with the limit requirements. Finally, the SOC at the end and at the beginning of the experiment were similar, about 40%, which ensured the adequacy of the batteries for the next experimental cycles.

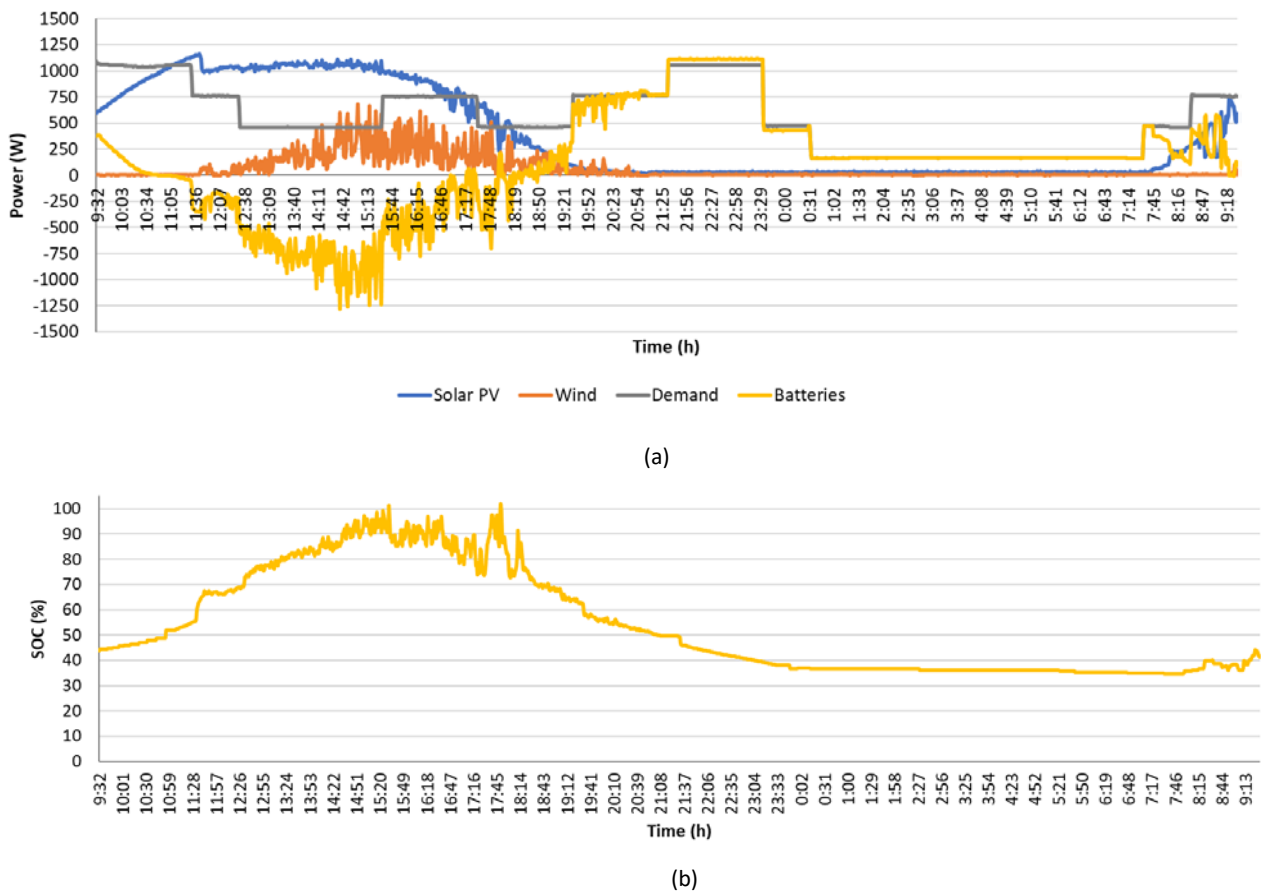

(a)

(b)

Figure 8. Experimental validation for the highest-scored configuration. (a) Energy Balance. (b) SOC.

Figure 9 plots the energy balance and SOC results for the second highest-scored configuration, which includes renewable generation and the support of batteries and a diesel generator.

The energy balance presented in this experiment is comparable to the previous one, with one main difference: the contribution of the diesel generator. The generator supplied energy during 1.5 h, at the beginning of the day. This option is very convenient because it guarantees the electricity supply during the period where the load demand is highest and solar irradiation and wind are still very low. Besides, the contribution of the diesel generator led to an increase of the batteries SOC from 35% to 85%.

The optimal use of the diesel generator demonstrated its suitability for the experiment: the rate of power losses was 4%, and the battery SOC at the end of the experiment (41%) was slightly higher than this value at the beginning of the experiment (35%), ensuring therefore the adequacy of the batteries for future energy cycles.

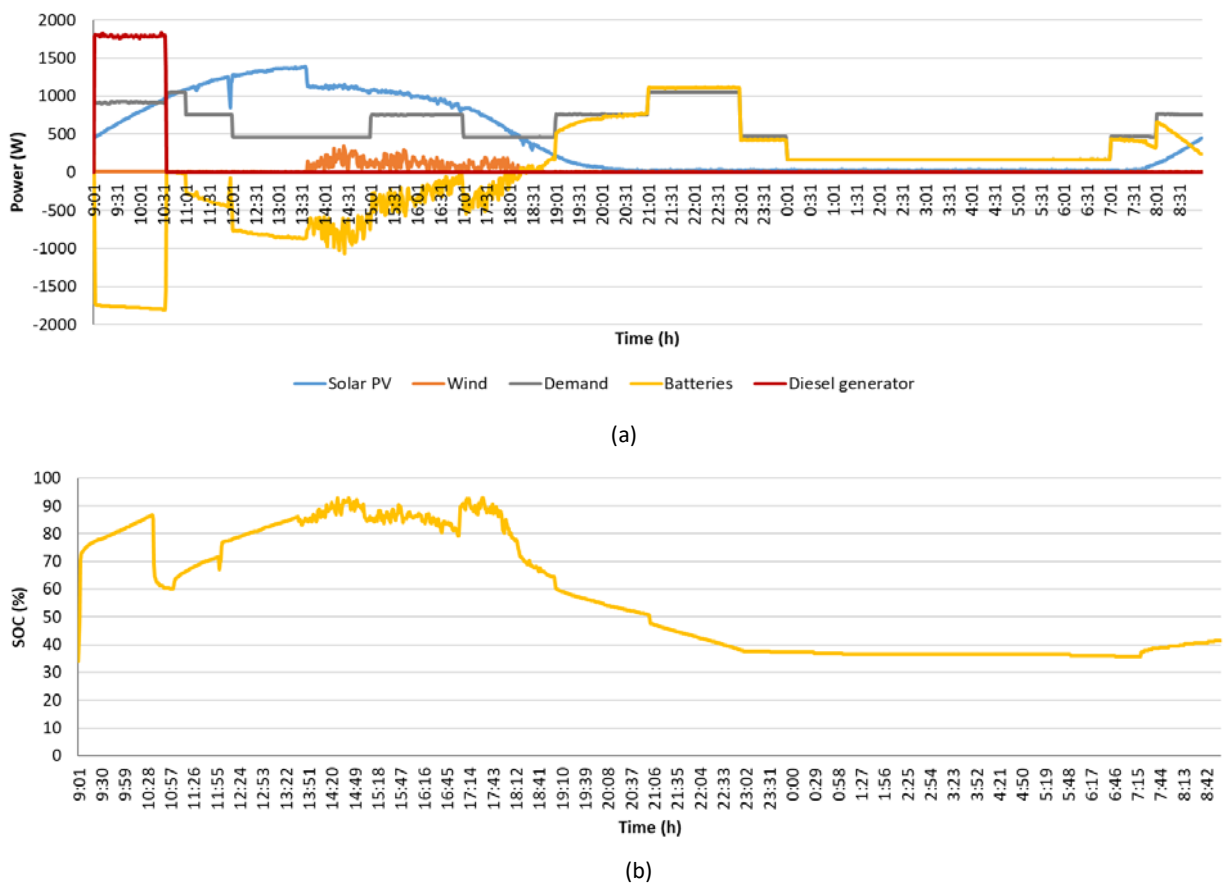

(a)

(b)

Figure 9. Experimental validation for the second highest-scored configuration. (a) Energy Balance. (b) SOC.

Figure 10 plots the energy balance results for the third highest-scored configuration, which includes renewable generation with the support of the grid.

As Figure 10 reflects, at early morning the grid covered the low solar irradiation at the period of maximum load demand. Later, there was an excess in generation from solar PV that was inyected into the grid. During this period, solar irradiation was available and wind contribution was higher than in the previous configuration checks. Hence, the grid was also responsible for absorbing the variability of the wind generation. Besides, the grid supplied the required electricity during the evening and night period. For this experiment, power losses acquired the value of 4%, meeting therefore the limit conditions.

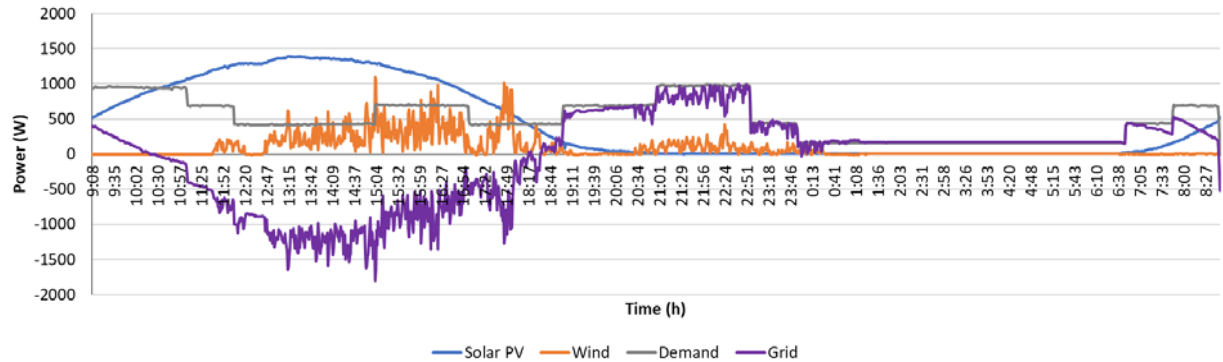

Figure 10. Experimental validation for the third highest-scored configuration. Energy Balance.

These experimental results demonstrated the energy balance suitability of the three selected configurations for the HRES in EVCS, both in the level of power losses and batteries' SOC limits and with a full time coverage of the load demand.

## 5. Conclusions

A high penetration of EVCS is expected to happen to cope with the electricity requirements of the also foreseeable high introduction of EVs in the medium-term future for almost all developing countries. This electrification of the transport sector arises as an environmental solution since EVs emit zero emissions while riding on the roads, although careful attention should be paid to the emissions in the generation of the electricity they need. The use of microgrids with renewable generation (HRES) in EVCS seems necessary, since this use would decrease both the CI content of the electricity generation and the pressure on the grid that the recharge of EVCS would produce. Choosing the most suitable configuration for HRES in EVCS, taking into account in its design the different constraints in the technical, economic and environmental aspects, would be an essential task.

This paper has defined a novel multicriteria methodology that takes into consideration all the above-mentioned constraints and includes an experimental stage to verify the configuration of the HRES for EVCS. The methodology, after the determination of the available renewable resources and the electricity demand of the EVCS, uses HOMER® code to deduce possible HRES configurations and evaluates them with a new multicriteria analysis, considering weighted technical, economic and environmental parameters to rank them. Finally, configurations with the highest scores are experimentally tested to check their reliability, power balance and SOC range. Hence, the selected final configuration design ensures the suitability of the HRES for the EVCS, supported not only by a complete numerical evaluation, but also by an experimental verification.

To illustrate the viability of the methodology, the article applies the method to the case study of Valencia, the capital province of Comunidad Valenciana, (in the east of Spain). This province is immersed in a remarkable mobility transition, with the aim of increasing the quantity of EVs and EVCS, together with a significant introduction of renewable sources in the electricity generation system.

Results for the electricity demand modelling of these vehicles in EVCS led to a maximum load demand of 270 kW that takes place during the early morning (from 9:00 to 10:00 h) and at early night again (from 21:00 to 22:00 h). On the other hand, the generation resources analysis revealed the suitability of solar PV and wind resources, with and average solar daily irradiation of 5 kWh/m$^2$/day and an average wind speed of 3.6 m/s at 18 m, respectively. Regarding back-up systems, batteries, diesel generator and grid connection were contemplated.

An initial simulation of the system considering both restrictions (generation resources availability and electricity demand) and making use of HOMER ® resulted in a starting filtered list of 27 configuration alternatives. These options were later evaluated by means of the hereby presented multicriteria methodology, with the same weights for the different constraints. Simulation results indicated that the most suitable configuration for the case study is an off-grid system with renewable generation and batteries support, followed by another off-grid system that includes also the support of a diesel generator. The third highest-scored configuration resulted in an on-grid system with renewable generation.

The selected configurations were experimentally validated in the Laboratory of Distributed Energy Resources (labDER) at the Polytechnic University of Valencia (Spain). Both the generation and demand resources were scaled according to the laboratory components with a factor of 1:250. Results indicated that the demand was fully covered in all the scenarios, with maximum power losses of 4.5% and SOC of batteries between 35% and 100%.

To conclude, this study provides a methodology that ensures the suitability of the HRES for the EVCS, supported not only by a complete multicriteria assessment, but also by an experimental verification. Its application to the case study of Valencia proves the viability of applying HRES for recharging EVs at EVCSs in a technical, economic and environmental acceptable way.

# 6. Acknowledgment

One of the authors (PBM) was supported by the regional public administration of Valencia under the grant ACIF/2018/106.

# 7. References

[1] Akitt JW. Some observations on the greenhouse effect at the Earth's surface. Spectrochim Acta Part A Mol Biomol Spectrosc 2018;188:127–34. https://doi.org/10.1016/J.SAA.2017.06.051.

[2] Dino IG, Meral Akgül C. Impact of climate change on the existing residential building stock in Turkey: An analysis on energy use, greenhouse gas emissions and occupant

comfort. Renew Energy 2019;141:828–46. https://doi.org/10.1016/j.renene.2019.03.150.

[3] Woo JR, Choi H, Ahn J. Well-to-wheel analysis of greenhouse gas emissions for electric vehicles based on electricity generation mix: A global perspective. Transp Res Part D Transp Environ 2017;51:340–50. https://doi.org/10.1016/j.trd.2017.01.005.

[4] Data & Statistics - IEA 2017. https://www.iea.org/data-and-statistics?country=WORLD&fuel=Energy consumption&indicator=Oil products final consumption by sector (accessed February 13, 2020).

[5] Teixeira ACR, Sodré JR. Impacts of replacement of engine powered vehicles by electric vehicles on energy consumption and CO2 emissions. Transp Res Part D Transp Environ 2018;59:375–84. https://doi.org/10.1016/J.TRD.2018.01.004.

[6] Dijk M, Orsato RJ, Kemp R. The emergence of an electric mobility trajectory. Energy Policy 2013;52:135–45. https://doi.org/10.1016/J.ENPOL.2012.04.024.

[7] Su J, Lie TT, Zamora R. Modelling of large-scale electric vehicles charging demand: A New Zealand case study. Electr Power Syst Res 2019;167:171–82. https://doi.org/10.1016/J.EPSR.2018.10.030.

[8] Liu Z, Wu Q, Nielsen A, Wang Y. Day-Ahead Energy Planning with 100% Electric Vehicle Penetration in the Nordic Region by 2050. Energies 2014;7:1733–49. https://doi.org/10.3390/en7031733.

[9] Manjunath A, Gross G. Towards a meaningful metric for the quantification of GHG emissions of electric vehicles (EVs). Energy Policy 2017;102:423–9. https://doi.org/10.1016/j.enpol.2016.12.003.

[10] Álvarez Fernández R. A more realistic approach to electric vehicle contribution to greenhouse gas emissions in the city. J Clean Prod 2018;172:949–59. https://doi.org/10.1016/j.jclepro.2017.10.158.

[11] Bastida-Molina P, Hurtado-Pérez E, Pérez-Navarro Á, Alfonso-Solar D. Light electric vehicle charging strategy for low impact on the grid. Environ Sci Pollut Res 2020:1–17. https://doi.org/10.1007/s11356-020-08901-2.

[12] Galiveeti HR, Goswami AK, Dev Choudhury NB. Impact of plug-in electric vehicles and distributed generation on reliability of distribution systems. Eng Sci Technol an Int J 2018;21:50–9. https://doi.org/10.1016/J.JESTCH.2018.01.005.

[13] Deb S, Tammi K, Kalita K, Mahanta P. Impact of Electric Vehicle Charging Station Load on Distribution Network. Energies 2018;11:178. https://doi.org/10.3390/en11010178.

[14] Dixon J, Bukhsh W, Edmunds C, Bell K. Scheduling electric vehicle charging to minimise carbon emissions and wind curtailment. Renew Energy 2020;161:1072–91. https://doi.org/10.1016/j.renene.2020.07.017.

[15] Wu C, Gao S, Liu Y, Song TE, Han H. A model predictive control approach in microgrid considering multi-uncertainty of electric vehicles. Renew Energy 2021;163:1385–96. https://doi.org/10.1016/j.renene.2020.08.137.

[16] Quddus MA, Kabli M, Marufuzzaman M. Modeling electric vehicle charging station expansion with an integration of renewable energy and Vehicle-to-Grid sources. Transp

Res Part E Logist Transp Rev 2019;128:251–79. https://doi.org/10.1016/j.tre.2019.06.006.

[17] Xie R, Wei W, Khodayar ME, Wang J, Mei S. Planning Fully Renewable Powered Charging Stations on Highways: A Data-Driven Robust Optimization Approach. IEEE Trans Transp Electrif 2018;4:817–30. https://doi.org/10.1109/TTE.2018.2849222.

[18] Huang P, Ma Z, Xiao L, Sun Y. Geographic Information System-assisted optimal design of renewable powered electric vehicle charging stations in high-density cities. Appl Energy 2019;255:113855. https://doi.org/10.1016/j.apenergy.2019.113855.

[19] Domínguez-Navarro JA, Dufo-López R, Yusta-Loyo JM, Artal-Sevil JS, Bernal-Agustín JL. Design of an electric vehicle fast-charging station with integration of renewable energy and storage systems. Int J Electr Power Energy Syst 2019;105:46–58. https://doi.org/10.1016/j.ijepes.2018.08.001.

[20] Chowdhury N, Hossain C, Longo M, Yaïci W. Optimization of Solar Energy System for the Electric Vehicle at University Campus in Dhaka, Bangladesh. Energies 2018;11:2433. https://doi.org/10.3390/en11092433.

[21] HOMER - Hybrid Renewable and Distributed Generation System Design Software 2020. https://www.homerenergy.com/ (accessed May 14, 2020).

[22] Vermaak HJ, Kusakana K. Design of a photovoltaic-wind charging station for small electric Tuk-tuk in D.R.Congo. Renew Energy 2014;67:40–5. https://doi.org/10.1016/j.renene.2013.11.019.

[23] Nizam M, Wicaksono FXR. Design and Optimization of Solar, Wind, and Distributed Energy Resource (DER) Hybrid Power Plant for Electric Vehicle (EV) Charging Station in Rural Area. Proceeding - 2018 5th Int. Conf. Electr. Veh. Technol. ICEVT 2018, Institute of Electrical and Electronics Engineers Inc.; 2019, p. 41–5. https://doi.org/10.1109/ICEVT.2018.8628341.

[24] Karmaker AK, Ahmed MR, Hossain MA, Sikder MM. Feasibility assessment & design of hybrid renewable energy based electric vehicle charging station in Bangladesh. Sustain Cities Soc 2018;39:189–202. https://doi.org/10.1016/j.scs.2018.02.035.

[25] Rashid MM, Islam Maruf MN, Akhtar T. An RES-based grid connected electric vehicle charging station for Bangladesh. 1st Int. Conf. Robot. Electr. Signal Process. Tech. ICREST 2019, Institute of Electrical and Electronics Engineers Inc.; 2019, p. 205–10. https://doi.org/10.1109/ICREST.2019.8644130.

[26] Tulpule PJ, Marano V, Yurkovich S, Rizzoni G. Economic and environmental impacts of a PV powered workplace parking garage charging station. Appl Energy 2013;108:323–32. https://doi.org/10.1016/j.apenergy.2013.02.068.

[27] Losev OG, Grigor'ev AS, Mel'nik DA, Grigor'ev SA. Charging Station for Electric Transport Based on Renewable Power Sources. Russ J Electrochem 2020;56:163–9. https://doi.org/10.1134/S1023193520020093.

[28] Savio DA, Juliet VA, Chokkalingam B, Padmanaban S, Holm-Nielsen JB, Blaabjerg F. Photovoltaic Integrated Hybrid Microgrid Structured Electric Vehicle Charging Station and Its Energy Management Approach. Energies 2019;12:168.

https://doi.org/10.3390/en12010168.

[29] Electric Mobility Plan 2017. https://www.gva.es/es/inicio/area_de_prensa/not_detalle_area_prensa?id=860077 (accessed July 2, 2020).

[30] Valencian Climate Change and Energy Strategy 2030 2017. http://www.agroambient.gva.es/es/web/cambio-climatico/2020-2030 (accessed July 2, 2020).

[31] Hansen JM, Xydis GA. Rural electrification in Kenya: a useful case for remote areas in sub-Saharan Africa. Energy Effic 2020;13:257–72. https://doi.org/10.1007/s12053-018-9756-z.

[32] Chowdhury T, Chowdhury H, Miskat MI, Chowdhury P, Sait SM, Thirugnanasambandam M, et al. Developing and evaluating a stand-alone hybrid energy system for Rohingya refugee community in Bangladesh. Energy 2020;191:116568. https://doi.org/10.1016/j.energy.2019.116568.

[33] Singh M, Balachandra P. Microhybrid Electricity System for Energy Access, Livelihoods, and Empowerment. Proc IEEE 2019;107:1995–2007. https://doi.org/10.1109/JPROC.2019.2910834.

[34] Driscoll Á, Lyons S, Mariuzzo F, Tol RSJ. Simulating demand for electric vehicles using revealed preference data. Energy Policy 2013;62:686–96. https://doi.org/10.1016/j.enpol.2013.07.061.

[35] Bastida-Molina P, Hurtado-Pérez E, Peñalvo-López E, Cristina Moros-Gómez M. Assessing transport emissions reduction while increasing electric vehicles and renewable generation levels. Transp Res Part D Transp Environ 2020;88:102560. https://doi.org/10.1016/j.trd.2020.102560.

[36] Hurtado E, Peñalvo-López E, Pérez-Navarro Á, Vargas C, Alfonso D. Optimization of a hybrid renewable system for high feasibility application in non-connected zones. Appl Energy 2015;155:308–14. https://doi.org/10.1016/J.APENERGY.2015.05.097.

[37] Xu X, Hu W, Cao D, Huang Q, Chen C, Chen Z. Optimized sizing of a standalone PV-wind-hydropower station with pumped-storage installation hybrid energy system. Renew Energy 2020;147:1418–31. https://doi.org/10.1016/j.renene.2019.09.099.

[38] Zhang Y, Yuan J, Zhao C, Lyu L. Can dispersed wind power take off in China: A technical & institutional economics analysis. J Clean Prod 2020;256:120475. https://doi.org/10.1016/j.jclepro.2020.120475.

[39] Hansen K. Decision-making based on energy costs: Comparing levelized cost of energy and energy system costs. Energy Strateg Rev 2019. https://doi.org/10.1016/j.esr.2019.02.003.

[40] Ribó-Pérez D, Bastida-Molina P, Gómez-Navarro T, Hurtado-Pérez E. Hybrid assessment for a hybrid microgrid: A novel methodology to critically analyse generation technologies for hybrid microgrids. Renew Energy 2020;157:874–87. https://doi.org/10.1016/j.renene.2020.05.095.

[41] Corporate Finance Institute. Levelized Cost of Electricity 2020.

https://corporatefinanceinstitute.com/resources/knowledge/finance/levelized-cost-of-energy-lcoe/ (accessed May 14, 2020).

[42] Hidalgo Batista ER, Villavicencio Proenza DD. The reliability of stationary internal combustion diesel engines. Rev Científica Trimest 2011:1–10.

[43] Kruyt B, van Vuuren DP, de Vries HJM, Groenenberg H. Indicators for energy security. Energy Policy 2009;37:2166–81. https://doi.org/10.1016/j.enpol.2009.02.006.

[44] Sovacool BK, Mukherjee I. Conceptualizing and measuring energy security: A synthesized approach. Energy 2011;36:5343–55. https://doi.org/10.1016/j.energy.2011.06.043.

[45] Bastida-Molina P, Hurtado-Pérez E, Vargas-Salgado C, Ribó-Pérez D. Hybrid micronetworks, a solution for developing countries. Técnica Ind 2020;325:28–34. https://doi.org/10.23800/10218.

[46] Pérez-Navarro A, Alfonso D, Ariza HE, Cárcel J, Correcher A, Escrivá-Escrivá G, et al. Experimental verification of hybrid renewable systems as feasible energy sources. Renew Energy 2016;86:384–91. https://doi.org/10.1016/J.RENENE.2015.08.030.

[47] DGT. Traffic information 2019. http://infocar.dgt.es/etraffic/ (accessed September 19, 2019).

[48] Zheng J, Sun X, Jia L, Zhou Y. Electric passenger vehicles sales and carbon dioxide emission reduction potential in China's leading markets. J Clean Prod 2020;243:118607. https://doi.org/10.1016/j.jclepro.2019.118607.

[49] Philipsen R, Brell T, Brost W, Eickels T, Ziefle M. Running on empty – Users' charging behavior of electric vehicles versus traditional refueling. Transp Res Part F Traffic Psychol Behav 2018;59:475–92. https://doi.org/10.1016/j.trf.2018.09.024.

[50] Martínez-Lao J, Montoya FG, Montoya MG, Manzano-Agugliaro F. Electric vehicles in Spain: An overview of charging systems. Renew Sustain Energy Rev 2017;77:970–83. https://doi.org/10.1016/J.RSER.2016.11.239.

[51] Sehar F, Pipattanasomporn M, Rahman S. Demand management to mitigate impacts of plug-in electric vehicle fast charge in buildings with renewables. Energy 2017;120:642–51. https://doi.org/10.1016/J.ENERGY.2016.11.118.

[52] Li J, Gao S, Xu B, Chen H. Modeling and Controllability Evaluation of EV Charging Facilities Changed from Gas Stations with Renewable Energy Sources. 2019 Asia Power Energy Eng. Conf. APEEC 2019, Institute of Electrical and Electronics Engineers Inc.; 2019, p. 269–73. https://doi.org/10.1109/APEEC.2019.8720700.

[53] Luca de Tena D, Pregger T. Impact of electric vehicles on a future renewable energy-based power system in Europe with a focus on Germany. Int J Energy Res 2018;42:2670–85. https://doi.org/10.1002/er.4056.

[54] REE. Electric mobility guide for local entities 2018. https://www.ree.es/sites/default/files/downloadable/Guia_movilidad_electrica_para_entidades_locales.pdf (accessed July 31, 2019).

[55] PVGIS 2020. http://re.jrc.ec.europa.eu/pvgis/apps4/pvest.php?lang=es&map=europe (accessed December 26, 2018).

[56] Wind resource analyses. Wind atlas of Spain 2019. https://www.idae.es/uploads/documentos/documentos_11227_e4_atlas_eolico_A_9b90ff10.pdf (accessed July 8, 2020).

[57] International Energy Agency. Data and statistics 2016. https://www.iea.org/data-and-statistics/data-tables?country=WORLD&energy=Balances&year=2016 (accessed December 12, 2019).

[58] Bastida Molina P, Saiz Jiménez JÁ, Molina Palomares MP, Álvarez Valenzuela B. Instalaciones solares fotovoltaicas de autoconsumo para pequeñas instalaciones. Aplicación a una nave industrial. 3C Tecnol 2017:1–14. https://doi.org/http://dx.doi.org/10.17993/3ctecno.2017.v6n1e21.1-14.